\documentclass[%
        preprint,preprintnumbers,%
        aps,prl,showpacs,amsmath,amssymb]{revtex4}
\usepackage{graphicx}

\newcommand{\ket}[1]{\mbox{$|#1\rangle$}}
\newcommand{\bra}[1]{\mbox{$\langle{}#1|$}}
\newcommand{\bracket}[2]{\langle #1 | #2 \rangle}
\newcommand{\braOket}[3]{\langle #1 | #2 | #3 \rangle}
\newcommand{\ketbra}[2]{| #1 \rangle \langle #2 |}

\newcommand{\set}[1]{\left\{ #1 \right\}}

\newcommand{\bbset}[1]{\mathbb{#1}}
\newcommand{\Integer}{\bbset{Z}}

\newcommand{\up}{\uparrow}
\newcommand{\down}{\downarrow}

\def\toexp{\mathop{\rm exp}}
\newcommand{\Texp}{\toexp_{\leftarrow}}

\newcommand{\Init}{{\rm I}}
\newcommand{\Fin}{{\rm F}}
\newcommand{\Ground}{0_{\rm B}}
\newcommand{\Answer}{{\rm Ans}}

\begin{document}

\title{Quasienergy anholonomy and its application to
  adiabatic quantum state manipulation}

\author{Atushi Tanaka}
\email{tanaka@phys.metro-u.ac.jp}
\affiliation{Department of Physics, Tokyo Metropolitan University,
  Minami-Osawa, Hachioji, Tokyo 192-0397, Japan}
\author{Manabu Miyamoto}
\email{miyamo@hep.phys.waseda.ac.jp}
\affiliation{Department of Physics, Waseda University, 
  Okubo, Shinjuku-ku, Tokyo 169-8555, Japan}


\begin{abstract}
The parametric dependence of a quantum map
under the influence of a rank-1 perturbation is investigated.
While the Floquet operator of the map and its spectrum have
a common period with respect to the perturbation strength $\lambda$,
we show an example in which none of the quasienergies nor the eigenvectors
obey the same period:
After a periodic increment of $\lambda$, 
the quasienergy arrives at the nearest higher one,
instead of the initial one, exhibiting an anholonomy, which
governs another anholonomy of the eigenvectors.
An application to quantum state manipulations is outlined.
\end{abstract}


\maketitle

Adiabaticity is substantial when a system in question separates 
from its ``environments,'' which are described by either almost 
``frozen'' degrees of freedom or slowly varying external parameters.
In quantum theory, the response of the bound states of the system
to infinitely slow change of the environment 
is one of the oldest subjects, and is summarized as the 
{\em adiabatic theorem}~\cite{Born:ZP-51-165}.
The change of a system's eigenenergy reflects just the exchange of energy, 
or the balance of the forces~\cite{HellmannFeynmanTheorem}
between the system and the environment.
The direction of the state vector follows that of an
instantaneous eigenvector, 
while the phase of the state vector is determined
by two different origins, a dynamical one that is associated 
with the eigenenergy, and 
a geometric one, which was overlooked for many years.
The latter phase is particularly prominent in the structure
of the eigenstate induced by ``global'' changes of the environment.
More precisely, after the environment slowly moves along a circuit
in the configuration space of the environmental parameter and returns to 
the initial point,
the phases of initial and final states of the system may not coincide, 
even when the dynamical phase is zero.
This discrepancy is called anholonomy 
(or holonomy, in terms of differential geometry)%
~\cite{Simon:PRL-51-2167,GeometricPhaseReview}.
A simple demonstration of the phase anholonomy is shown 
by Berry~\cite{Berry:PRSLA-392-45}.
Subsequently, a ``non-Abelian'' extension of phase anholonomy 
in the presence of degenerate eigenenergies is pointed out 
by Wilczek and Zee, where the geometric phase factor 
is non-commutative~\cite{Wilczek:PRL-52-2111}.
The phase anholonomy appears in various fields of physics, besides
quantum mechanics, and brings profound 
consequences~\cite{GeometricPhaseReview}.

While there have been many studies on the phase anholonomy,
an anholonomy in eigenvalues has been recognized only recently in physics:
Cheon discovered an anholonomy in eigenenergies, in a family of 
systems with generalized pointlike potentials~\cite{Cheon:PLA-248-285}:
The trail of an eigenenergy along a change of parameters 
on a closed path that encircles a singularity does not draw a closed curve 
but, instead, a spiral. 
The anholonomy induces another anholonomy in the directions of eigenvectors:
The adiabatic changes of the parameter along the closed path cause a state 
vector that is initially prepared in an eigenvector of the Hamiltonian
to travel to another eigenspace corresponding to a different eigenenergy
even without any degenerate eigenenergies.
This is completely different from Wilczek-Zee's phase anholonomy,
which needs a degenerate, multi-dimensional eigenspace 
in which the state vector can rotate.
In order to distinguish the anholonomies in the phase and the direction of
an eigenvector, we call the latter an {\em eigenspace} anholonomy.
The origin of the eigenvalue and eigenspace anholonomies 
in the family of systems with generalized pointlike potentials, 
is elucidated in terms of
the geometrical structure of the system's parameter space%
~\cite{Cheon:AP-294-1}.
Up to now, the examples of Cheon's anholonomies in
physical systems are few, and their realization seems to 
require a singular potential~\cite{Tsutui:JPA-36-275}.

In spite of their uncommonness, it is still true that 
Cheon's anholonomies touch upon the very fundamental point
of adiabaticity in quantum theory. Thus they would present us with a
tremendous number of implications and applications, as
the phase anholonomy does.
For example, if Cheon's anholonomies are experimentally accessible, 
one easily expects 
an important application to
be adiabatic manipulations of quantum states~\cite{Cheon:PLA-248-285},
which we will discuss in the latter part of this paper.
Cheon's anholonomies may enable us to realize the most primitive
adiabatic control on the population of an adiabatic state that
is almost classical.
Accordingly, this control would be far more robust 
than the controls that rely on constructive quantum interferences.
It is worth pointing out that the applications of adiabatic processes to 
the control of quantum states have already become textbook results%
~\cite{QuantumControlTextbook}. At the same time, quantum circuits 
and computers using the phase anholonomy are expected to be robust 
due to their 
topological nature~\cite{Zanardi:PLA-264-94,Jones:Nature-403-869}.

The first aim of this paper is to show Cheon's anholonomies 
in quantum maps.
More precisely, we will discuss anholonomies both in quasienergies
and in eigenspaces
of Floquet operators that describe unit time 
evolutions of the quantum maps~\cite{QuantumMap}.
Our example works with a Floquet operator with a discrete spectrum, 
under 
a rank-1 perturbation~\cite{Combescure:JSP-59-679,endnote:TermPerturbation}. 
This means that we have a systematic way 
to produce instances of quasienergy and associated eigenspace anholonomies.
Hence we may argue that Cheon's anholonomies are abundant in 
systems that are described by quantum maps.
The second aim is to demonstrate an application of the quasienergy and 
eigenspace anholonomies to manipulation of quantum states, which is
straightforward at least theoretically.
As an example, we outline an implementation of an anholonomic adiabatic 
quantum computation.

Our minimal example is a two-level system, whose unperturbed
Hamiltonian is $\hat{H}_0 = \frac{1}{2}\pi\hat{\sigma}_z$.
We set $\hbar = 1$ throughout this paper.
With a periodically 
pulsed rank-$1$ perturbation $\hat{V}=\ket{v}\bra{v}$, where
$\ket{v}$ is normalized, the system is described by a
``kicked'' Hamiltonian
$\hat{H}(t) \equiv
\hat{H}_0 + \lambda \hat{V}\sum_{n=-\infty}^{\infty}\delta(t - nT)$,
where $\lambda$ and $T$ are the strength and the period of 
the perturbation, respectively. 
We focus on the stroboscopic time evolutions of 
the state vector $\ket{\psi_n}$ just before the kick at $t = n T$.
The corresponding quantum map is 
$\ket{\psi_{n+1}} = \hat{U}_{\lambda}\ket{\psi_n}$,
where 
$\hat{U}_{\lambda} \equiv
\lim_{\epsilon\downarrow 0}
\Texp\left(-i\int_{-\epsilon}^{T-\epsilon} \hat{H}(t) dt\right)
= e^{-i\hat{H}_0 T} e^{-i\lambda\hat{V}}$
is a Floquet operator, and 
$\Texp$ is the time-ordered exponential~\cite{QuantumMap}.
We examine the eigenvalues $z_n(\lambda)$ ($n = 0, 1$) of $\hat{U}_{\lambda}$,
and the corresponding normalized eigenvector 
$\ket{\xi_n(\lambda)}$~\cite{endnote:EigenstateOfFloquetOp}.
The unitarity of $\hat{U}_{\lambda}$ ensures that the quasienergy 
$E_n(\lambda)\equiv i T^{-1}\ln z_n(\lambda)$ takes a real value, which is an 
``average of the energy'' (with modulo $2\pi T^{-1}$)
during the unit time interval.

In order to simplify the following argument, we introduce two assumptions:
(i) The spectrum of $\hat{U}_{0}$ is 
nondegenerate~\cite{endnote:DegenerateU0};
and (ii) $\ket{\xi_n(0)}$ is not any eigenvector of $\hat{V}$.
The latter implies 
$0 < |\bracket{v}{\xi_n(\lambda)}| < 1$ for all $\lambda$ and $n$, 
due to the fact that either $|\bracket{v}{\xi_n(\lambda)}| = 0$ or $1$ 
for some $\lambda$ contradicts with 
the assumption~\cite{endnote:OverlapVandXi}.

We explain a topological structure of the parameter space of $\lambda$
for $\hat{U}_{\lambda}$.
Since $\hat{V}$ is a projection operator,
$\hat{U}_{\lambda} = \hat{U}_{0} \{1 - (1 - e^{-i\lambda}) \hat{V}\}$ 
is periodic in $\lambda$ with period $2\pi$~\cite{Combescure:JSP-59-679}.
Hence the parameter space of $\lambda$ can be regarded 
as a circle $S^1$.
The periodicity of $\hat{U}_{\lambda}$ about $\lambda$ implies
that the quasienergy spectrum $\set{E_0(\lambda), E_1(\lambda)}$ also 
obeys the same periodicity.
Namely, after the increment of $\lambda$ by $2\pi$, i.e., a ``cycle'' in
the parameter space $S^1$, both  $\hat{U}_{\lambda}$ and
the set $\set{E_0(\lambda), E_1(\lambda)}$ return to the initial points.

\begin{figure}
  \includegraphics[width=0.40\textwidth]{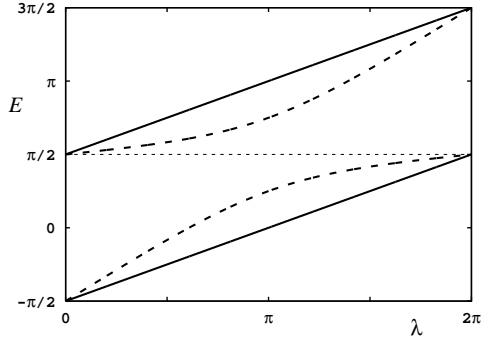}
  \caption{\label{fig:twolevel}
    Parametric motions of quasienergies of two-level 
    model systems explained in the main text, 
    with the period of the time interval $T=1$.
    The branch of the quasienergy is chosen as $[-\pi/2, 3\pi/2)$.
    The bold parallel lines correspond to the case where
    $\ket{v}=(\ket{\up}-i\ket{\down})/\sqrt{2}$.
    The quasienergies are $(\lambda \pm \pi)/2$.
    The corresponding eigenvectors 
    $\ket{\xi_+(\lambda)} = 
    \cos(\lambda/4)\ket\up + \sin(\lambda/4)\ket\down$
    and 
    $\ket{\xi_-(\lambda)} = 
    -\sin(\lambda/4)\ket\up + \cos(\lambda/4)\ket\down$
    also exhibit eigenspace anholonomy.
    Namely, $\ket{\xi_s(0)}$ and $\ket{\xi_s(2\pi)}$ 
    are orthogonal ($s=\pm$).
    The bold-dashed curves, which have an avoided crossing,
    correspond to the case where
    $\ket{v} = \cos(\pi/8)\ket{\up} + \sin(\pi/8)\ket{\down}$.
    Note that the minimal gap between two curves
    depends on the ratio $|\bracket{\up}{v}|:|\bracket{\down}{v}|$.
  }
\end{figure}

After establishing the periodicity of $\lambda\in S^1$,
we now examine each quasienergy to seek an anholonomy.
First of all, the branch of quasienergies is chosen as
$[E_0(0), E_0(0)+2\pi T^{-1})$. 
Because of the nondegeneracy of $\hat{U}_0$, 
we have $E_0(0) < E_1(0) < E_0(0)+2\pi T^{-1}$.
To examine how much $E_n(\lambda)$ increases during a cycle 
of $\lambda$, we evaluate
$\Delta E_n \equiv 
\int_0^{2\pi}\left(\partial_{\lambda} {E_n(\lambda)}\right) d\lambda$,
where
$\partial_\lambda {E_n(\lambda)}$
is the rate of the change of $n$-th quasienergy
against the change of $\lambda$.
Note that $\Delta E_n$ is ``quantized''
due to the periodicity of the spectrum,
e.g., $\Delta E_0$ is either $0$ or $E_1(0) - E_0(0) \mod 2\pi T^{-1}$.
To determine which is the case, 
we evaluate the integral expression of  $\Delta E_n$ 
with 
$\partial_\lambda {E_n(\lambda)}
= T^{-1} \braOket{\xi_n(\lambda)}{\hat{V}}{\xi_n(\lambda)}$~%
\cite{Nakamura:PRA-35-5294}.
Since the eigenvalues of $\hat{V}$ are $0$ and $1$, 
we have $0 \le \partial_\lambda {E_n(\lambda)} 
\left(= T^{-1} |\bracket{v}{\xi_n(\lambda)}|^2\right) \le T^{-1}$.
However, the equalities for the minimum and the maximum do not hold,
because $0 < |\bracket{v}{\xi_n(\lambda)}| < 1$, as stated above.
Hence, we have $0 < \Delta E_n < 2\pi T^{-1}$. 
Because of the quantization of $\Delta E_n$, 
we conclude $\Delta E_0 = E_1(0) - E_0(0)$, which
assures $E_0(\lambda)$ converges to $E_1(0)$ as 
$\lambda\nearrow 2\pi$.
Thus it is shown that the quasienergies $E_n(\lambda)$ as well as 
the eigenvalues $z_n(\lambda)$
do not return to the initial values at $\lambda=0$ after 
the parameter goes around
a cycle of the parameter space (see Fig.~\ref{fig:twolevel}). 
This is nothing but a manifestation of Cheon's anholonomy in quasienergy.

An extension of the example above to $N$-level systems, with 
the assumptions (i) and (ii), also provide a similar example of 
the quasienergy anholonomy. This time, each $E_n(0)$ is transported to 
$E_{n+1\mod N}(0)$ after a cycle of $\lambda$. 
%
This is understood from the fact that all increment $ \Delta E_n $ 
during a cycle of $\lambda$ should satisfy the sum rule 
$\sum_{n=0}^{N-1}\Delta E_n 
= \int_0^{2\pi} T^{-1} ({\rm Tr}\hat{V})d\lambda
=2\pi T^{-1}$.
In fact, as in the two-level cases, each $ E_n (\lambda) $ should increase 
and finally reach $ E_{n + \nu \mod N} (0) $ ($1 \le \nu < N$) 
as $\lambda\nearrow 2\pi$. 
However, if $\nu > 1$, the total increment 
$\sum_{n=0}^{N-1}\Delta E_n $ results in breaking the sum rule stated above.
Thus only $\nu = 1$ is allowed for all $n$.

The quasienergy anholonomy induces another non-conventional anholonomy 
in eigenvectors, i.e., the eigenspace anholonomy.
Let us consider an adiabatic transport of the eigenvector 
$\ket{\xi_n(\lambda)}$ of $\hat{U}_{\lambda}$ following the slow 
changes of $\lambda$ with an asymptotically long step $M$($\gg 1$).
Then, the state vector evolves as
$\hat{U}_{\lambda_{M-1}}\hat{U}_{\lambda_{M-2}}\ldots
\hat{U}_{\lambda_{0}}\ket{\xi_n(\lambda_0)}$, where 
$\lambda_m$ denotes the value of $\lambda$ at the $m$-th step.
According to the adiabatic theorem for eigenvectors of Floquet 
operators~\cite{Holthaus:PRL-69-1596}, the state vector 
stays in an instantaneous eigenvector of $\hat{U}_{\lambda}$
continuously if the change of $\lambda$ is 
slow enough~\cite{endnote:AdiabaticTheoremForQuantumMap}.
In our case, when an eigenvector of $\hat{U}_{\lambda}$
is adiabatically transported along a cycle of $\lambda$, the resultant
eigenvector is orthogonal to the initial one (see caption in 
Fig.~\ref{fig:twolevel}). This is because the eigenvectors
corresponding to different eigenvalues are orthogonal to each other,
due to the unitarity of $\hat{U}_{\lambda}$.

Our geometrical interpretation of the quasienergy anholonomy resembles
Cheon's eigenenergy anholonomy in the generalized pointlike 
potentials~\cite{Cheon:AP-294-1}. 
This is natural because the families of both models are parameterized by
$2$-dimensional unitary matrices.
We may employ a space of two quasienergies
$\{(E_0, E_1)\}$ as a parameter space of the Floquet operators of 
two-level systems, 
with a suitable identification such that
an element $(E_0, E_1)$ is identified with 
$(E_1, E_0)$. 
The quotient quasienergy-spectrum-space 
is accordingly
an orbifold $T^2/\Integer_2$, which has two topologically
inequivalent and nontrivial cycles (see, Ref.~\cite{Cheon:AP-294-1}). 
One cycle crosses the degeneracy line $E_0 = E_1$.
The other cycle transports the quasienergy
from $E_0(0)$ to $E_1(0)$. The increment of $\lambda$ in $\hat{U}_{\lambda}$
actually follows the latter cycle. 
The geometrical nature suggests that the quasienergy anholonomy is
stable against perturbations that preserve the topology of the cycle.
Hence we may expect that the same anholonomy appears in
other than periodically kicked systems, e.g., periodically driven systems.

In the following, we discuss applications of Cheon's
anholonomies in quantum maps to the manipulations of quantum states.
As is shown above, 
it is possible to convert a state vector, which is initially in
an eigenstate of nondegenerate Floquet operator $\hat{U}_0$, 
to the nearest higher eigenstate 
of $\hat{U}_0$, by applying a periodically pulsed perturbation 
$\hat{V}=\ketbra{v}{v}$, 
whose strength $\lambda$ is adiabatically increased from 0 to $2\pi$,
as long as $\ket{v}$ satisfies the condition mentioned above.
Note that at the final stage of the control, we may switch off
the perturbation suddenly, due to the periodicity of the 
Floquet operator under the rank-$1$ perturbation
$\hat{U}_{2\pi} = \hat{U}_{0}$.
This closes a ``cycle.''
By repeating the cycle, the final state can be 
an arbitrary eigenstate of $\hat{U}_0$. 
As a control scheme, the initial and final states of our procedure 
are only the eigenstates of the Floquet operator and not their superpositions.
The advantage of our procedure is the following:
(1) This is widely applicable: as long as
the spectrum of $\hat{U}_0$ contains only discrete components, 
we can work with it.
(2) The scheme is robust, thanks to the adiabaticity. 
In particular, $\ket{v}$ is allowed to vary adiabatically.
Namely, slow fluctuations on $\ket{v}$ do not harm controls.
At the same time, our scheme is not influenced by the presence of
dynamical phases~\cite{endnote:RemarkOnQCwithPhaseAnholonomy}.

In order to demonstrate the potential applicability of the anholonomic 
quantum state manipulations with Cheon's anholonomies,
we explain an idea of anholonomic adiabatic quantum computation,
which is an anholonomic variant of 
Farhi {\it et al\/}'s adiabatic quantum computation~%
\cite{AdiabaticComputer}.
Before describing our approach, we explain the conventional procedure.
Its aim is to find a solution, expressed by a number $n$, of a problem P,
which is composed by conditions on the solution $n$.
In the following, we assume that P has only a single solution.
For example, when P is the 3-satisfiability problem (3-SAT) 
of $N$-bit numbers,
the cost of finding a solution of P is generally $\mathcal{O}(2^N)$
as $N\to\infty$, 
i.e., exponentially difficult~\cite{HopcroftSAT}.
The following ``Hamiltonian formulation'' provides a way to
solve P with the help of quantum theory.
Let $H_{\rm P}(n)$ be a ``cost function,'' or, a ``Hamiltonian,'' of P,
indicating 
the number of conditions that are not satisfied
by a number $n$. 
The ground state of ${H}_{\rm P}(n)$, i.e., the value of $n$
that satisfies ${H}_{\rm P}(n) = 0$, is the solution of P.
In order to describe the ``arithmetic register'' $n$ with quantum theory,
we introduce a basis $\set{\ket{n}}$.
Accordingly, the quantized Hamiltonian is
$\hat{H}_{\rm P} = \sum_n\ket{n}{H}_{\rm P}(n)\bra{n}$~%
\cite{endnote:DefHP}.
Now the procedure to find the answer of P is mapped to
a problem in obtaining the ground state $\ket\Answer$, of $\hat{H}_{\rm P}$.
To solve P, Farhi {\it et al}. proposed employing
the adiabatic theorem~\cite{AdiabaticComputer}:
Let us start from an initial Hamiltonian  $\hat{H}_{\rm B}$,
whose ground state is well known $\ket{0_{\rm B}}$, and assume
that the ground energy is $0$.
For example, we may employ
$\hat{H}_{\rm B} = 
\beta (\hat{1} - \ket{\Ground}\bra{\Ground})$, where $\beta$ is
positive~\cite{Znidaric:PRA-73-022329}.
To use the adiabatic theorem, an interpolation Hamiltonian 
$\hat{H}(t)\equiv 
(1 - t/T_{\rm r}) \hat{H}_{\rm B} + (t/T_{\rm r}) \hat{H}_{\rm P}$
is introduced, where $T_{\rm r}$ is the ``running time.''
At $t=0$, the state of the arithmetic register is prepared 
to be in  $\ket{0_{\rm B}}$,
and the state will arrive at the ground state of $\hat{H}_{\rm P}$ when
$t = T_{\rm r}$, if $T_{\rm r}$ is large enough to ensure the adiabatic
condition, which is determined by the energy gap between the ground state
and the first excited state of $\hat{H}(t)$.
Some numerical experiments on 3-SAT show that $T_{\rm r}$
grows only polynomially as a function of the system size $N$,
while it is proven that $T_{\rm r}$ grows exponentially, i.e., inefficiently, 
with the specific choice 
of $\hat{H}_{\rm B}$ shown above~\cite{Znidaric:PRA-73-022329}. 
To overcome this inefficiency, 
there seems to be room for further investigations, for example,
to proper choices of 
the initial Hamiltonian $\hat{H}_{\rm B}$~\cite{endnote:AckToReferee},
or the intermediate adiabatic process~\cite{Farhi:quant-ph-0208135}.
Our strategy also might provide a workaround, as is suggested below.

To explain our anholonomic adiabatic quantum computation,
we reuse the Hamiltonians $\hat{H}_{\rm B}$ and $\hat{H}_{\rm P}$
and the arithmetic register of the conventional adiabatic
quantum processor.
An additional qubit is employed as a ``control register,''
whose Hilbert space is spanned by 
orthonormal vectors $\ket\Init$ and $\ket\Fin$, which indicate
the initial and the final states of the computation, respectively.
Next we introduce an ``unperturbed Hamiltonian'' 
$
  \hat{H}_0 \equiv
  (\hat{H}_{\rm B} - \epsilon) \otimes \ketbra{\Init}{\Init} 
  + \hat{H}_{\rm P} \otimes \ketbra{\Fin}{\Fin},
$
where $0 < \epsilon < \beta$ is assumed.
Then, $\hat{H}_0$ has the following properties:
(1) The ground state $\ket\Ground\otimes\ket\Init$ is nondegenerate and
the ground energy $-\epsilon$ is negative.
(2) Because $\beta - \epsilon > 0$, 
the eigenenergy of the first excited state 
$\ket\Answer\otimes\ket\Fin$ is $0$ and nondegenerate too.
These two ``target'' states are also eigenstates of 
the unperturbed Floquet operator $\hat{U}_0\equiv e^{-i\hat{H}_0 T}$.
To ensure that there is no quasienergy of $\hat{U}_0$, between
the quasienergies of the two target states,
it is sufficient to choose 
the period of the kicks to satisfy $T < 2\pi / W$, 
where $W$ is the difference between the maximum and the minimum 
eigenenergy of $\hat{H}_0$.
Therefore, imposing a periodically kicked rank-$1$ perturbation 
$\hat{V}=\ket{v}\bra{v}$, 
the state vector, which is initially prepared to be
$\ket\Ground\otimes\ket\Init$,
is adiabatically transported to $\ket{\Answer}\otimes\ket{\Fin}$ 
with the help of the quasienergy anholonomy of
$\hat{U}_{\lambda} = \hat{U}_0 e^{-i\lambda\hat{V}}$,
where $\ket{v}$ needs to have non-zero overlap with each target state.
We remark that the degeneracies of other quasienergies of $\hat{U}_0$
do not detract from our purpose~\cite{endnote:DegenerateU0}.
To achieve an efficient computation, the quasienergy gaps
around the ``ground state''
need to be large enough during the adiabatic transport.
The present scheme might offer a way to prevent the disastrous slowdown 
of the running time
with a proper adjustment of $\ket{v}$.
For example, if we take into account only the two target states approximately, 
the gap becomes maximum if we use $\ket{v}\propto 
\ket{\Ground}\otimes\ket{\Init} + \ket{\Answer}\otimes\ket{\Fin}$
(see Fig.~\ref{fig:twolevel}).
This choice, however, would be impossible 
unless we know $\ket{\Answer}$.
Namely, there needs to be a compromise on the choice of $\ket{v}$
in order to realize our scheme with reasonable resources.
We leave this point as an open question, 
which must be clarified to evaluate the efficiency 
of the present approach~\cite{endnote:Equivalence}.
As a final remark, the simplicity of the above proposal, one of 
the largest scale applications of the anholonomic quantum state 
manipulations, indicates that Cheon's anholonomies 
in quantum maps deserve further investigations.


\begin{acknowledgments}
  M.M. would like to thank Professor I.~Ohba and
  Professor H.~Nakazato for useful comments.
  This work is supported in part (M.M.) 
  by a grant for the 21st-Century COE Program at Waseda 
  University from 
  MEXT, Japan.
\end{acknowledgments}


\begin{thebibliography}{28}
\bibitem{endnote:TermPerturbation}
  We remark that the strength of our ``perturbation'' need not to be small
  and that we do not rely on any kind of series expansions 
  with small parameters.
\bibitem{endnote:EigenstateOfFloquetOp}
  Corresponding to an eigenvector $\ket{\xi_n(\lambda)}$, the periodic 
  time-dependence of state vector except its dynamical phase between kicks 
  is described by a Floquet state~\cite{QuantumMap,Holthaus:PRL-69-1596}.
\bibitem{endnote:DegenerateU0}
  Degenerate spectrums will be treated in Ref~\cite{TM060}.
\bibitem{endnote:OverlapVandXi}
  If we assume $\bracket{v}{\xi_n(\lambda)} = 0$, 
  $\ket{\xi_n(\lambda)}$ is a simultaneous eigenvector of $\hat{V}$
  and $\hat{U}_0$, due to the fact
  $\hat{U}_0\ket{\xi_n(\lambda)} 
  = \hat{U}_{\lambda}e^{+i\lambda\hat{V}}\ket{\xi_n(\lambda)}
  = z_n(\lambda)\ket{\xi_n(\lambda)}$.
  %
  Similarly, under the assumption $|\bracket{v}{\xi_n(\lambda)}| = 1$, 
  $\ket{\xi_n(\lambda)}$ is an eigenvector of 
  both $\hat{V}$ and $\hat{U}_0$.
\bibitem{endnote:RemarkOnQCwithPhaseAnholonomy}
  In applications of the phase anholonomy to quantum circuits,
  it is required to remove the influences from dynamical 
  phases~\cite{Jones:Nature-403-869}.
\bibitem{endnote:DefHP}
  An efficient implementation of $\hat{H}_{\rm P}$ is shown in 
  Ref.~\cite{AdiabaticComputer}.
\end{thebibliography}



\end{document}